\documentclass[a4paper]{tweppart}

\title{Production and Testing of the LHCb Outer Tracker Front End Readout Electronics}
\author{E.~Simioni \\[14pt]
On behalf of the LHCb Outer Tracker group\\[15pt]
NIKHEF, The Netherlands\\[14pt]
\texttt{eduard.simioni@cern.ch}
\vspace{1cm}
}
\begin{document}
\maketitle
\begin{multicols}{2}
\begin{abstract} 
The LHCb Outer Tracker is a straw drift detector with a modular design and a total of
53760 readout channels distributed over a sensitive area of 12 double layers of 6x5
$m^{2}$ each.
The main electronics readout requirement is the precise (~0.5 ns) drift time
measurement at an occupancy of $\sim$4$\%$ and 1 MHz readout. 
A total of 128 channels are read out by one Front-End box.
About half of the FE$-$Boxes have been built. Quality Assurance during the production has been
performed on single FE$-$Box components.
The assembled FE$-$Box is finally commissioned using a special FE-Tester.
The FE$-$Tester is a programmable pulser with a time resolution of 150 ps capable to
simulate all the functionality of the readout mimicking the real detector.
Consequently, problems have been found and solved resulting in good
overall performance.
\end{abstract}
\section{Overview of the Readout System}
The FE$-$Box readout has a modular structure consisting of four
different type of boards: HV board (decoupling the analog
signal from the high voltage), ASDBLR aboard (amplification), OTIS board (drift time measurement) and GOL 
auxiliary board (supply voltage and the optical link for data transmission). 
The boards are installed in an aluminum frame built to fit on the straw module providing the mechanical 
protection and electrical shielding to the electronics inside.
The analog part of the electronics chain is the most critical and it will be treated with more emphasis.
\subsection{Amplifier}
\begin{figure}
\begin{center}
  \includegraphics[width=90mm]{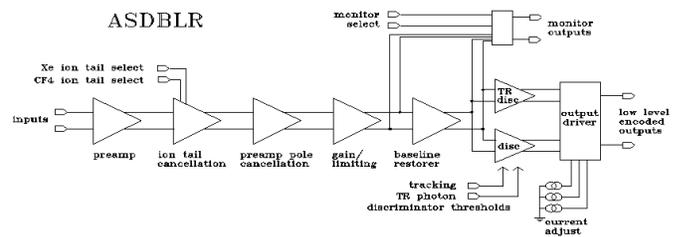}
\end{center}
\caption{Block diagram of the ASDBLR Chip.}
\label{asdblr}
\end{figure}

Charge signals from the straw detector are decoupled from the high voltage and 
amplified, shaped and discriminated by the ATLAS ASDBLR chips. 
The main characteristics of the ASDBLR chip are:
\begin{enumerate}
\item Eight input channels and two test pulses (low and high).
\item A fast peaking time of $\sim$6ns due to a shaping network and the baseline restoration
which separate the leading edge of the input signal and cut the ions induced current tail. 
\item Radiation hardness withstanding $3.5$ $10^{14}~ n/cm^{2}$ \cite{asd_radhard}.
\item Low cross talk of ~0.2$\%$ and low noise of $<$1fC equivalent charge thanks
to the DMILL bipolar process adopted for the chip production.
\item Optimized for grounding and heat dissipation through a 100 $\mu$m
copper layer endowed in the PCB and several vias between the board layers.
\end{enumerate}
The block diagrams of the chip functionality is shown in Fig.\ref{asdblr}.
Two ASDBLR chip are assembled on one ASDBLR board.
\subsection{Time to Digital Converter (TDC)}
Digitized signal from two ASDBLR boards are then collected by the OTIS board \cite{otis_b} (Outer tracker Time Information
System) which contains a 32 channel TDC. The chip is developed 
using a standard 0.25 $\mu$ CMOS process.
The drift time data of 
each channel is stored in a pipeline of 4 $\mu$s waiting for the trigger veto.
%
%
\subsection{Gigabit Optical Link (GOL)}
At L0 accept the output of the TDC is collected by one Gigabit Optical Link chip (GOL)
assembled in the GOL auxiliary board. The GOL board collects the signals coming from four OTIS boards
and low voltage supply ($+$2.5 V and $\pm$ 3 V), TFC signals (Time and Fast Control)
and $I^{2}$C slow control bus. Optical fibers carry the data $\sim$90 meters far
from the detector at the L0 output rate of 1.1 MHz to the TELL1 board \cite{tell1} 
in order to be filtered and finally stored for off-line processing.
\subsection{FE$-$Box}
\begin{figure}
\begin{center}
  \includegraphics[width=80mm]{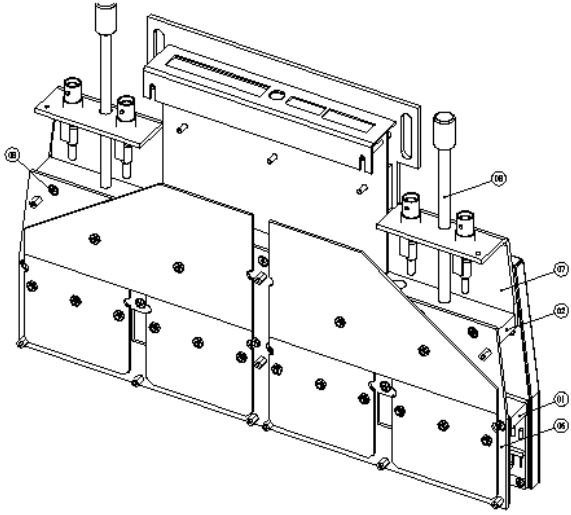}
\end{center}
\caption{Outline of the FE$-$Box chassis with boards.}
\label{chassis}
\end{figure}

The FE$-$Box consists of an aluminum chassis designed to fit one end of the detector module 
on which are installed 4 HV boards, 8 ASDBLR boards, 4 OTIS boards and 1 GOL auxiliary board. 
The FE$-$Chassis provides the cooling system, four inputs for the HV, the shielding cover and
two connectors to fix the FE$-$Box on the straw module. In Fig.\ref{chassis} the drawing of as 
open FE$-$Box is shown.
\section{FE$-$Test Setup Description}
The electronics readout requirements are a precise ($\sim$0.5 ns) and efficient
drift time measurement at an occupancy of $~$4$\%$ to ensure
single hit resolution (200 micron) and efficient charged particle reconstruction.
To achieve the desired performance, several steps of quality assurance
during the production have been applied, using dedicated test setups
for each type of board \cite{test_boa}.\\
\begin{figure}
\begin{center}
  \includegraphics[width=90mm]{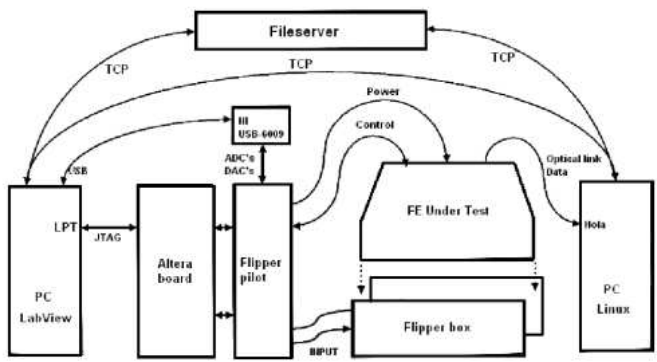}
\end{center}
\caption{Block diagram of the FE Test Setup.}
\label{setup}
\end{figure}
The performance on an assembled FE$-$Box is verified through 
a final test performed using a special FE$-$Tester. The block diagram of the setup
is shown in Fig.\ref{setup}.
The FE$-$Tester is based on the test setup build for the Alice Alcapone tester.
The heart of the setup is a PCB with an Altera programmable logic chip.
Most of the electronics needed for the tests is built on the controller board.
To interface the FE$-$Box a specific connection board is developed
(Flipper Box) with the additional required electronics to provide the input signal on the
128 channels of the FE$-$Box.
The logic in the Altera chip is controlled by a LabView program on a PC. The
connection between them is a JTAG interface through the parallel port.
For the communication with the FE$-$Box the I2C bus is used.
The out-coming data stream from the GOL board is collected by the HOLA acquisition
board \cite{hola}. The data exchange between the two machine and a file-server is
guarantee by standard TCP.\\
In summary the FE$-$Tester consists of a programmable pulser
capable to provide all the functionality of the readout (slow and fast controls) 
mimicking the input signals as real module detector. the FE$-$Test functionality 
are here summarized:
\begin{enumerate}
\item Generation of input signals straw-like. Those signals are tunable in 
intensity and in time with a resolution of $\sim$0.5 fC and 150 ps respectively.
\item Generation TFT (Time and Fast Control).
\item Generation of I2C (Slow Control).
\item Power supply and data acquisition.
\end{enumerate}
All those operations are automatized and controlled through LabView.
%
%
%
\section{Test of the FE$-$Box}
The following tests have been performed on assembled FE$-$Boxes:
\begin{enumerate}
\item Threshold Characteristics. A threshold scan is done and the measurement of the
half-efficiency-point is carried out for a fixed input charge.
The relative variation of the half-efficiency-point is expected to be
less than 60 mV (rougly corresponding to half fC) out of the 128 channels as required in the ASDBLR chip selection \cite{asd_sel}.
A threshold scan is also performed using the test pulse signals (low and high) generated in the ASDBLR
chip. The last test on the preamplifier is done through an amplitude scan with fixed threshold to carry out 
the half-efficiency-point and the ENC (equivalent noise charge).
\item Timing. Measurement of the time conversion linearity and the channel-by-channel
resolution of the OTIS board.
\item Noise. Dark noise is studied as function of the threshold. 
\item Synchronization. Four 8 bit TDC chips are inside a single FE$-$Box. The time
difference between test pulse and the L0 is checked with a latency scan
and therefore the synchronization between channels is verified.\\
\end{enumerate}
\subsection{Threshold Characteristics}
One general requirement of the preamplifier, is the channel-by-channel uniformity response for the same
input charge $Q_{i}$. The efficiency, noise and occupancy are strongly sensitive to the threshold
which influence the detector performance. The threshold must be chosen such that the uniformity between channels
is guaranteed in a wide range of input charge.\\
Given the function $g(Q_{i})$ representing the signal amplitude after pre-amplification of the input charge $Q_{i}$,
to measure channel-by-channel uniformity we need to "compare" those amplitudes between channels defining
an appropriate amplitude estimator. 
The easiest number to measure is the hit-efficiency.
Varying the threshold from low to high, the hit-efficiency goes from 1 to 0.
In the ideal condition of absence of noise the hit-efficiency profile against threshold would be modeled as a step function. 
Assuming Gaussian noise 
the expression for the hit-probability for a fixed threshold and input charge is \cite{lhcb117}:
\begin{equation}
Pr(V_{thr},Q_{i}) = N \int_{V_{thr}}^{+\infty} e^{-\frac{1}{2}~\big(\frac{V-g(Q_{i})}{\sqrt{2}~\sigma_{noise}}\big)^2}dV
\label{eq1}
\end{equation}
Where $N$ is a normalization factor, $V$ is the threshold and $\sigma_{noise}$ is the Gaussian noise amplitude.
The Eq. \ref{eq1} can be rewritten using the standard erf function:
\begin{equation}
Pr(V_{thr},Q_{i}) = \frac{1}{2}~-~\frac{1}{2}Erf \Big( \frac{V_{thr}-g(Q_{i})}{\sqrt{2}~\sigma_{noise}} \Big)
\label{eq2}
\end{equation}
\begin{figure}
\begin{center}
  \includegraphics[width=55mm]{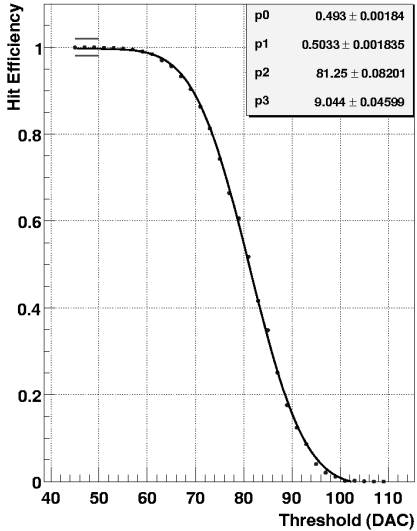}
\end{center}
\caption{Hit-efficiency as function of threshold for a fixed input charge.}
\label{thr_eff}
\end{figure}
The hit-efficiency is 50$\%$. for $V_{thr}=g(Q_{i})$.
The threshold in which the condition above is realized is called half-efficiency threshold
and from now on we refer to it as $V_{thr}^{50\%}$, which is the estimator for $g(Q_{i})$.
By measuring the hit-efficiency profile against threshold and fitting the Erf function model, we can determine channel-by-channel
the $V_{thr}^{50\%}$ and the $\sigma_{noise}$ as shown in Fig \ref{thr_eff}. The hypotesis of gaussian error fits the data points.
\subsubsection{Requirements in the $V_{thr}^{50\%}$ uniformity}
Each ASDBLR chip has been individually tested at UPENN university before assembly. 
By the test results, chips have been first pre-selected in order to separate chips with working problems 
and then further selected applying constraints on the $V_{thr}^{50\%}$ uniformity. 
The two main parameter defined for a single ASDBLR chip are:
\begin{enumerate}
\item $\Delta V_{thr}^{max}$ : The absolute maximum deviation by the average in the half-efficiency-point measured on the
single ASDBLR (8 channels).
\item $\Delta V_{thr}^{G~max}$ : The absolute maximum deviation by the global average in the half-efficiency-point
(measured on the total amount of preselected chips).
\end{enumerate}
Five different values of injected charge have been used in the chip test, (0, 3, 5, 30 and 50 fC) and
the cuts applied after pre-selection are summarized in Table \ref{tab1}.
\begin{table}
\begin{center}
\begin{tabular}{|c|c|c|}
\hline
Parameter &  Charge & Cut \\
\hline
$\Delta V_{thr}^{max}$ & 5 fC  & $\pm$ 30 mV \\
\hline
$\Delta V_{thr}^{max}$ & 30 fC & $\pm$ 60 mV \\
\hline
$\Delta V_{thr}^{Gmax}$ & 5 fC & $\pm$ 30 mV \\
\hline
\end{tabular}
\caption{Selection cuts on $V_{thr}^{50\%}$ uniformity for ASDBLR chip.}
\label{tab1}
\end{center}
\end{table}
\subsubsection{Threshold Scan with the FE$-$Setup}
The channel-by-channel $V_{thr}^{50\%}$ is determined 
in three ways: using the FE$-$Box input and the two test-pulses in the ASDBLR chip.
\begin{enumerate}
\item From the input: injection of the signals through the flipper-box (straw-like signals).
\item From Test-pulse: signals generated by the slow-control and routed through the GOL board and the four OTIS until the 
the 16 ASDBLR ICs. In the ASDBLR there are two test-pulse inputs: high test-pulse and low test-pulse. The amplitudes 
of those test-pulses are not tunable.
\end{enumerate}
The functioning and the performance of the ASDBLR once assembled on the PCB is guaranteed by three 
different channel-by-channel determinations of the $V_{thr}^{50\%}$.\\
About 2$\%$ of assembled ASDBLR chips show a channel broken or having a large deviations. Usually this is due to problems
in the chip soldering, missing components or damaged ASDBLR board input$/$output connector pins.
The uniformity in the $V_{thr}^{50\%}$ is in good agreement with the requirements adopted in the 
chip selection.
%
%
%
\subsubsection{Systematics in the input Threshold Scan}
The $V_{thr}^{[50\%]}$ is determined with good accuracy of $\sim$1$\%$ and systematic deviations due to the setup (in particular
the flipper box) are detectable. As shown in Fig. \ref{setup}, the flipper box receives from the control pilot two signals, one for each
side of the FE$-$Box. Each signal is then split through 64 channels but the amount of charge available slowly decreases going from the first 
to the last channel.
Since the half-efficiency-point is a function of the injected charge the first channel has higher value of $V_{thr}^{[50\%]}$.\\ 
%
%
This systematic can be estimated and an offset to each channel is added.
An offset up to the $\sim4\%$ has been measured, 
based on a Gaussian fit to the
128 distributions of $V_{thr}^{[50\%]}\left(\mathrm{FE,ch}\right)$
measured in testing the FE$-$Boxes. The values of the means
$\left\{ \mu\left(\mathrm{ch}\right): \mathrm{ch}=1, ..., 128 \right\}$
from the Gaussian fits are then used to calculate the offsets:
\begin{equation}
        \epsilon(\mathrm{ch}) \equiv \frac{1}{128}
        \sum_{k=1}^{128}\mu(k)-\mu(\mathrm{ch}),
        \label{offset}
\end{equation}
which are used to correct the values of $V_{thr}^{[50\%]}$ measured:
\begin{equation}
        V_{thr}^{[50\%]}\left(\mathrm{FE,ch}\right)
        \rightarrow
        V_{thr}^{[50\%]}\left(\mathrm{FE,ch}\right) + \epsilon(\mathrm{ch}).
        \label{eff_off}
\end{equation}
%
%
\subsubsection{Half-efficiency-point in the input charge amplitude scan}
\begin{figure}
\begin{center}
  \includegraphics[width=55mm]{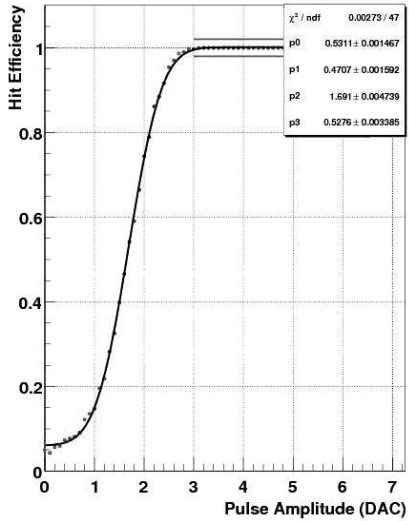}
\end{center}
\caption{Hit-efficiency as function of amplitude charge for a fixed threshold.}
\label{chareff}
\end{figure}

As shown in Fig. \ref{chareff}, increasing the input charge for a fixed threshold the hit efficiency 
goes from 0 to 1. 
As in the threshold scan, the ideal profile would be a step function but
assuming Gaussian noise the following model can be used to fit the data points:
\begin{equation}
Pr(Q_{i},V_{thr}) = \frac{1}{2}~+~\frac{1}{2}Erf \Big( \frac{g(Q_{i})-V_{thr}}{\sqrt{2}~\sigma_{noise}} \Big)
\end{equation}
\label{eq2}
Now assuming that amplifier response is linear \cite{lhcb117} we can write:
\begin{equation}
Pr(Q_{i},V_{thr}) = \frac{1}{2}~+~\frac{1}{2}Erf \Big( \frac{g(Q_{i})-Q_{thr}}{\sqrt{2}~ENC} \Big)
\end{equation}
\label{eq2}
Obtaining a direct measurement of the ENC (equivalent noise charge).\\
This measurement is strongly anti-correlated and complementary to the one performed in the threshold scan.
\subsection{Timing}
The aim of the timing test is to measure the channel by channel linearity and resolution.
\subsubsection{Linearity of OTIS}
\begin{figure}
\begin{center}
  \includegraphics[width=55mm]{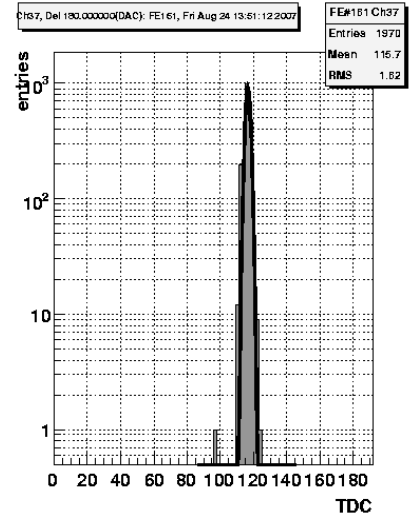}
\end{center}
\caption{Example of drift time distribution with a Gaussian fit.}
\label{drift_time}
\end{figure}

The FE$-$Setup provides a straw-like signal with high time accuracy (0.15 ns). 
Tuning the delay between the L0 and the input signal (for fixed threshold and input charge) a delay scan is performed.
Considering the overall steps and matching the mean measured drift time (see Fig. \ref{drift_time}) with 
the time delay set, the linear correlation shown in Fig. \ref{linearity} is obtained for each channel. The straight line
runs over the readout window of 192 TDC (75 ns).\\
One out of 20 FE$-$Boxes shows bad linearity or large offset deviant from other channels.
\subsubsection{Systematics in the input Delay Scan}
\begin{figure}
\begin{center}
  \includegraphics[width=55mm]{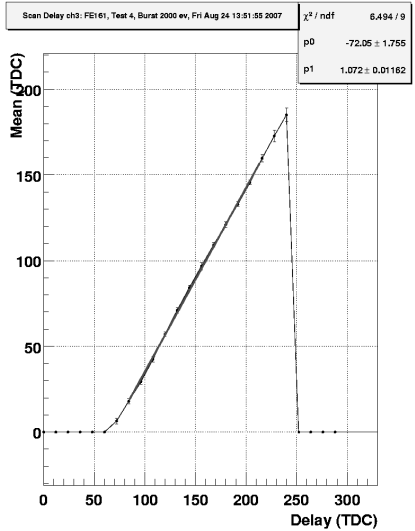}
\end{center}
\caption{Linearity of the OTIS board.}
\label{linearity}
\end{figure}

Due to the high time accuracy, systematics due to the setup are detectable. The flipper pilot generates two input signals
which goes through the flipper box as explained in the previous section. 
The time offset between the two main signals is easily detectable with the oscilloscope and can
be corrected tuning the cable length or introducing a delay unit. 
A second source of systematics is due to the fact that each main signal is split
between 64 channels. As conseguence the signal has to go through the flipper box yielding a channel-by-channel delay
of about $\sim$0.2 ns. This time offset does not affect the linearity studies but only shift the straight line on Fig. \ref{linearity}.
\subsubsection{Resolution}
By the sigma of the Gaussian fit applied on the drift time spectra, the single-channel time resolution is estimated.
Noisy channels or channels with double-peak in the drift spectra are detected by with a large value of sigma. 
Combining all the drift time spectra collected in the first part of the production we can infer the 
average time resolution of the OTIS: 1.3 $\pm$ 0.2 TDC which correspond to 0.50 $\pm$ 0.06 ns.
%
%
%
\subsection{Noise}
A threshold scan is performed without input. For very low threshold (20 DAC, corresponding to $\sim$200 mV) the hit efficiency 
is 1 but increasing the threshold it decreases
exponentially to zero. This dependence is studied for each channel aiming to find noisy channels or problems due to
bad shielding cover. About one out of 30 FE$-$Box are repaired due to this.
\subsection{Synchronization}
Drift time information per event is stored in a 4 $\mu$s pipeline buffer in the OTIS chip waiting for the LO trigger decision.
Since four OTIS chips are embedded in a FE$-$Box it is crucial to verify whether the latency is equal for all the channels.
Varying the delay between the high Test-Pulse and the L0, a latency scan in steps of 12.5 ns is performed to verify
the overall synchronization across the read-out windows of 75 ns.
\section{Summary}
The FE$-$Setup has been used during the R$\&$D phase of the Outer Tracker electronics. 
Its usefulness was proven by finding problems at an early stage and finally it was crucial 
to improve the overall performance. Secondly, the FE$-$Tester was used during mass 
production of the FE$-$Boxes, and commissioned each FE$-$Box before delivery to 
CERN for the installation on the Outer Tracker.
\end{multicols}
\end{document}